\documentclass[aps,pre,showpacs,
amssymb,twocolumn
]{revtex4-2}
\usepackage{graphicx}
\usepackage{amsmath}
\usepackage{amsfonts}
\usepackage{amssymb}
\usepackage{epsfig,libertine}
\usepackage[libertine]{newtxmath}
\usepackage{color}
\usepackage{dsfont, hyperref, xcolor}
\usepackage{comment}
\usepackage{enumerate}

\usepackage{amsthm} %%%proof env

\newcommand{\numberset}{\mathbb}

\newcommand{\R}{\numberset{R}}

\newcommand{\Tr}[1]{\text{Tr}\left\{#1\right\}}

\newcommand{\bra}[1]{\langle#1\vert}
\newcommand{\ket}[1]{\vert#1\rangle}
\newcommand\braket[2]{\langle#1|#2\rangle}

\begin{document}

\title{Work fluctuation theorems with initial quantum coherence}

\author{Gianluca~Francica and Luca Dell'Anna}
\address{Dipartimento di Fisica e Astronomia e Sezione INFN, Università di Padova, via Marzolo 8, 35131 Padova, Italy}

\date{\today}

\begin{abstract}
Fluctuation theorems are fundamental results in nonequilibrium thermodynamics beyond the linear response regime. Among these, the paradigmatic Tasaki-Crooks fluctuation theorem relates the statistics of the works done in a forward out-of-equilibrium quantum process and in a corresponding backward one. In particular, the initial states of the two processes are thermal states and thus incoherent in the energy basis. Here, we aim to investigate the role of initial quantum coherence in work fluctuation theorems, by considering a quasiprobability distribution of work. To do this, we formulate and examine the implications of a detailed fluctuation theorem, which reproduces the Tasaki-Crooks fluctuation theorem in the absence of initial quantum coherence.
\end{abstract}

\maketitle
\section{Introduction}
Work is a fundamental nonequilibrium stochastic quantity, which plays a relevant role in out-of-equilibrium processes generated by changing some parameters of the system in a certain time interval. In classical systems, when there is equilibrium at the initial time, the work fluctuations satisfy an integral fluctuation relation, the Jarzynski equality~\cite{jarzynski97}. Furthermore, the statistics of the work done in the process is related to its time reversal by a detailed fluctuation relation given by the Crooks fluctuation theorem~\cite{crooks99}, when both the forward and backward processes start from equilibrium states. In particular, the integral fluctuation relation can be obtained from the detailed one by integrating it. When the quantum effects cannot be neglected, the detailed fluctuation relation still holds if we describe the statistics of work with the help of a two-projective-measurement scheme, as it was originally shown in Ref.~\cite{talkner07}, after the relevant works in Refs.~\cite{kurchan00,tasaki00}. This is known as Tasaki-Crooks fluctuation theorem (see, e.g., Ref.~\cite{campisi11} for a review).
However, in the presence of initial quantum coherence in the energy basis, in this scheme, where two projective measurements of the energy are performed at the initial and final times to infer the work statistics, the first measurement destroys the initial quantum coherence in the energy basis, and the protocol becomes invasive since irreversibly changes the real process.

There are several attempts to describe the work statistics in quantum regime (see, e.g., Refs.~\cite{deffner16,talkner16,allahverdyan14,solinas15,miller17,sampaio18,anza22,cerisola23,lostaglio23}). If we require that the two-projective-measurement statistics is also reproduced for incoherent initial states, a no-go theorem~\cite{perarnau-llobet17} suggests that the statistics of work can be represented by a quasiprobability distribution. Furthermore, if some conditions are satisfied, the work can be described by a class of quasiprobability distributions~\cite{Francica22,Francica222}, which includes the ones of Refs.~\cite{allahverdyan14,solinas15}.
The initial states of the Tasaki-Crooks fluctuation theorem are incoherent mixtures of the energy basis, and so are incoherent states. Then, the effects coming from the initial quantum coherence in the energy basis, e.g., quantum contextuality, are absent in this case. For instance, the effects of the initial quantum coherence can be taken in account by using conditional probabilities evaluated via Bayes' rule~\cite{rodrigues23}. However, in this case the probability distribution achieved is not linear with respect to the initial state, which is a fundamental requirement from our point of view. Thus, to preserve linearity, we will allow the distribution to take negative values obtaining a quasiprobability.
 In general, it is not clear how any of the established fluctuation theorems change when the system is prepared in
non-equilibrium states, e.g., recently this problem was also investigated by considering specific systems (see, e.g., Ref.~\cite{Yadalam22}).
However, differently from this and several other investigations concerning quantum coherence and energy exchange (see, e.g., Refs.~\cite{Ansari15,Uzdin16,Ansari19} to name a few), in the present paper we base our analysis on a quasiprobability approach.

Here, we aim to investigate the effects of the initial quantum coherence in the work fluctuation theorems in general. After introducing some preliminary notions in Sec.~\ref{sec. prelimi}, we discuss the time reversal of the quasiprobability distribution in Sec.~\ref{sec. time rev}. In particular, although the forward process can have a non-contextual representation, its time reversal can exhibits contextuality. Thus, we derive our main result in Sec.~\ref{sec. deta},  a detailed fluctuation theorem which holds in the presence of initial quantum coherence. In detail, it reproduces the Tasaki-Crooks fluctuation relation in the absence of quantum coherence and implies two different integral fluctuation theorems (see Sec.~\ref{sec. inte}), one of which was introduced in Ref.~\cite{Francica22}. To do this, we will focus on initial states with thermal populations and nonzero coherence in the energy basis, such as the coherent Gibbs state. %While we leave the quantum coherence of the forward initial state arbitrary, we fix the quantum coherence of the initial state of the backward process.

\section{Preliminaries}\label{sec. prelimi}
We start our discussion by introducing some preliminary notions, which are the Tasaki-Crooks fluctuation theorem (see Sec.~\ref{sec. incohe}), the quasiprobability distribution of work (see Sec.~\ref{sec. quasi}) and the quantum contextuality (see Sec.~\ref{sec. contex}).

\subsection{Incoherent initial state: Tasaki-Crooks fluctuation theorem}\label{sec. incohe}
We focus on a quantum coherent process generated by changing some parameters of the system in the time interval $[0,\tau]$. Thus, we get the time-dependent Hamiltonian $H(t)=\sum \epsilon_k(t) \ket{\epsilon_k(t)}\bra{\epsilon_k(t)}$, where $\ket{\epsilon_k(t)}$ is the eigenstate with eigenvalue $\epsilon_k(t)$ at the time $t\in [0,\tau]$, which generates the unitary time evolution operator $U_{t,0}=\mathcal T e^{-i\int_0^t H(s) ds}$, where $\mathcal T$ is the time order operator. The system is prepared at the initial time $t=0$ in an initial state $\rho$, which evolves to the final state $\rho'=U_{\tau,0}\rho U_{\tau,0}^\dagger$ at the time  $t=\tau$.
There is initial quantum coherence in the energy basis if there are non-zero coherences (i.e., off-diagonal elements of the density matrix) with respect to the energy basis. However, before discussing the effects of the initial quantum coherence, in this section we recall some results concerning an incoherent initial state, such that $\rho = \Delta(\rho)$, where we have defined the dephasing map $\Delta(\rho) = \sum_i \Pi_i\rho\Pi_i$, with the initial projectors $\Pi_i=\ket{\epsilon_i}\bra{\epsilon_i}$ and $\epsilon_i=\epsilon_i(0)$.
For an incoherent initial state $\rho$ the work can be represented by the two-projective-measurement scheme~\cite{talkner07,campisi11}, which in general has the probability distribution
\begin{equation}\label{eq. TPM}
 p_{\text{TPM}}(w) = \sum_{k,i} \Tr{\Pi_i\rho \Pi_i \Pi'_k} \delta(w-\epsilon'_k + \epsilon_i)\,,
\end{equation}
%so that the average work reads $\langle w\rangle = \int w p(w) dw$,
where the final projectors are defined as $\Pi'_k = U_{\tau,0}^\dagger\ket{\epsilon'_k}\bra{\epsilon'_k}U_{\tau,0} $ and $\epsilon'_k = \epsilon_k(\tau)$. % (in particular, if the spectrum of $H(0)$ is degenerate, we choose the basis of eigenstates $\ket{\epsilon_j}$ such that the restriction of $\rho_0$ on any eigenspace is diagonal with respect to such basis).
%We note that $\Tr{\Pi_i\rho \Pi_i \Pi'_k}=\Tr{\rho \Pi_i \Pi'_k}=\bra{\epsilon_i}\rho\ket{\epsilon_i} \abs{\bra{\epsilon'_k}U_{\tau,0}\ket{\epsilon_i}}^2 $.

The backward process is obtained by considering the backward time evolution from $\tau$ to zero, given by the unitary operator $U_{\tau-t,\tau}=U^\dagger_{\tau,\tau-t}$, where $t$ goes from zero to $\tau$. This backward process will start from an initial state $\bar \rho$. The two-projective-measurement scheme leads to the probability distribution for the backward process
\begin{equation}
 \bar{p}_{\text{TPM}}(w) = \sum_{k,i} \Tr{\bar{\Pi}'_k\bar{\rho} \bar{\Pi}'_k \bar{\Pi}_i} \delta(w- \epsilon_i+\epsilon'_k)\,,
\end{equation}
where in general we define $\bar{\Pi}'_k = U_{\tau,0} \Pi'_kU^\dagger_{\tau,0} =\ket{\epsilon'_k}\bra{\epsilon'_k}$ and $\bar{\Pi}_i = U_{\tau,0} \Pi_iU^\dagger_{\tau,0} =U_{\tau,0}\ket{\epsilon_i}\bra{\epsilon_i}U^\dagger_{\tau,0}$. If the initial state of the forward process is a Gibbs state at a certain inverse temperature $\beta$, i.e., $\rho=\rho_\beta\equiv e^{-\beta H(0)}/Z$, where $Z=\Tr{e^{-\beta H(0)}}$, for the forward process we get the Jarzynski equality~\cite{jarzynski97,talkner07,campisi11}
\begin{equation}\label{eq. Jarz}
\langle  e^{-\beta(w-\Delta F)}\rangle = 1\,,
\end{equation}
which has been also experimentally verified (see, e.g., Ref.~\cite{batalhao15}),
where the equilibrium free energy difference reads $\Delta F = -\beta^{-1}\ln(Z'/Z)$, where  $Z'=\Tr{e^{-\beta H(\tau)}}$.
In this case, if the initial state $\bar \rho$ of the backward process is the Gibbs state $\bar\rho=\rho'_\beta\equiv e^{-\beta H(\tau)}/Z'$, we get the Tasaki-Crooks fluctuation relation~\cite{tasaki00,talkner07,campisi11}
\begin{equation}\label{eq. FT crooks}
 e^{-\beta(w-\Delta F)} p_{\text{TPM}}(w) = \bar{p}_{\text{TPM}}(-w)\,.
\end{equation}
In particular, by integrating Eq.~\eqref{eq. FT crooks} over $w$ we achieve the integral fluctuation theorem of Eq.~\eqref{eq. Jarz}.
Our main aim is to generalize the detailed fluctuation theorem of Eq.~\eqref{eq. FT crooks} in the presence of initial quantum coherence. Since the two-projective-measurement scheme erases the initial quantum coherence, Eq.~\eqref{eq. FT crooks} still holds for any $\rho$ and $\bar{\rho}$ such that $\Delta (\rho)=\rho_\beta$ and $\bar\Delta(\bar\rho)=\rho'_\beta$, where $\bar{\Delta}(\bar{\rho}) = \sum_k \bar{\Pi}'_k\bar{\rho}\bar{\Pi}'_k$, and thus also in the presence of initial quantum coherence.
However, the two-projective-measurement work does not satisfy the first law of thermodynamics (see, e.g., Ref.~\cite{perarnau-llobet17}) and in this case the work can be represented by a quasiprobability distribution.

\subsection{Quasiprobability distribution of work}\label{sec. quasi}
Given a quantum observable $W$, having the spectral decomposition $W=\sum w_n P_n$, the projective measurements with projectors $\{P_n\}$ lead to the probability distribution $p_{\text{obs}}(w)=\sum_n v(P_n)\delta(w-w_n)$, where $v(P_n)=\Tr{P_n \rho}$ is a probability given by the Born rule, for the density matrix $\rho$. In general, for an effect $E$, which plays the role of event, the probability $v(E)=\Tr{E\rho}$ results from the Gleason's theorem (see, e.g., Ref.~\cite{bush03}). Here, we are considering an out-of-equilibrium process generated by changing some parameters in the time interval $[0,\tau]$. The first law of thermodynamics leads to the average work
\begin{equation}\label{eq. work1}
\langle w\rangle = \Tr{(H^{(H)}(\tau) -H(0) )\rho}\,,
\end{equation}
where $\rho$ is the initial density matrix and given an operator $A(t)$ we define the Heisenberg time evolved operator $A^{(H)}(t) = U_{t,0}^\dagger A(t) U_{t,0}$. Although $W=H^{(H)}(\tau) -H(0)$ is a quantum observable, in general its statistics does not reproduce the two-projective-measurement statistics for a Gibbs initial state $\rho=\rho_\beta$, so that the Jarzynski equality in Eq.~\eqref{eq. Jarz} is not satisfied (as originally observed in Refs.~\cite{Yukawa00,Allahverdyan05}). Actually, we can have two noncommuting quantum observables defining the work, which are $H(0)$ and $H^{(H)}(\tau)$. They give two sets of projectors $\{\Pi_i\}$ and $\{\Pi'_k\}$. In this case, Gleason’s theorem cannot be used in order to achieve a distribution of work that is linear with respect to the initial state. However, we can generalize Gleason’s axioms so that they lead to quasiprobabilities, e.g., $v(E,F)=\text{Re}\Tr{EF \rho}$, instead of the probabilities $v(E)$ (see, e.g., Appendix for details).

%\section{Time reversal}
%We introduce the projectors $\Pi_i = \ket{\epsilon_i}\bra{\epsilon_i}$ and $\Pi'_k=U_{\tau,0}^\dagger \ket{\epsilon'_k}\bra{\epsilon'_k} U_{\tau,0}$, which will represent our events, where $\epsilon_i=\epsilon_i(0)$ and $\epsilon'_k=\epsilon_k(\tau)$.
From these quasiprobabilities, we can write a quasiprobability distribution. In general, the work will be represented in terms of the events $\Pi_{i}$, $\Pi_j$, $\cdots$, $\Pi'_k$, $\cdots$, and thus we consider a quasiprobability distribution of the form~\cite{Francica222}
\begin{equation}\label{eq. p}
p(w) = \sum_{i,j,\ldots} v(\Pi_i,\Pi_j ,\ldots, \Pi'_k,\ldots ) \delta(w-w(\epsilon_i,\ldots))\,,
\end{equation}
%where the quasiprobability $v(\Pi_i,\Pi_j, \ldots, \Pi'_k,\ldots )$ is linear in the initial density matrix $\rho$, e.g., $v(\Pi_i,\Pi'_k ) = \text{Re}\Tr{\Pi_i\Pi'_k\rho}$ if there are only two events, which are $\Pi_i$ and $\Pi'_k$ (e.g., see Appendix for further details).
where the support is given by $w(\epsilon_i,\ldots)$, which is a function of the eigenvalues of the initial and final Hamiltonian.
The moments of the work are $\langle w^n\rangle = \int w^n p(w) dw$, where $n$ is an integer.
Without loss of generality, for the quasiprobability $v(\Pi_i,\Pi_j ,\ldots, \Pi'_k,\ldots )$ we can focus on definite decompositions of the proposition $\Pi_i\land\Pi_j\land\cdots$, since an affine combination of these quasiprobabilities gives any $p(w)$ of the form in Eq.~\eqref{eq. p}.
We note that in this case $v(\Pi_i,\Pi_j ,\ldots, \Pi'_k,\ldots )$ is the real part of the extended Kirkwood-Dirac quasiprobability of Refs.~\cite{YungerHalpern18,YungerHalpern24}.
As shown in Ref.~\cite{Francica222}, by requiring that (W1) the quasiprobability distribution $p(w)$ reproduces the two-projective-measurement statistics, i.e., $p(w)=p_{\text{TPM}}(w)$ when the initial state is incoherent in the energy basis, i.e., $\rho=\Delta(\rho)$, (W2) the average work is equal to Eq.~\eqref{eq. work1} and (W3) the second moment of work is equal to
\begin{equation}\label{eq. work2}
\langle w^2\rangle = \Tr{(H^{(H)}(\tau) -H(0) )^2\rho}\,,
\end{equation}
we get a class of quasiprobability distributions of the form~\cite{Francica22,Francica222}
\begin{equation}\label{eq. pq}
p_q(w) = \sum_{i,j,k} \text{Re}\Tr{\Pi_i \rho \Pi_j \Pi'_k} \delta(w-\epsilon'_k + q \epsilon_i + (1-q)\epsilon_j)\,,
\end{equation}
with $q$ real number. We note that the three conditions (W1), (W2) and (W3) define the support of the quasiprobability distribution, which otherwise is arbitrary without them and thus can represent anything. In particular, if (W3) is satisfied together with (W1) and (W2), the support linearly depends on the initial and final energy levels, leading to reasonable physical properties of the work (see Ref.~\cite{Francica222}).

\subsection{Contextuality} \label{sec. contex}
Knowing the quasiprobability distribution, we can focus on the problem if there is a non-contextual hidden variables model which satisfies the conditions about the reproduction of the two-projective-measurement scheme (W1), the average (W2) and the second moment (W3).
To introduce the concept of contextuality at an operational level~\cite{Spekkens05,Spekkens15,lostaglio18}, we consider a set of preparations procedures $P$ and measurements procedures $M$ with outcomes $k$, so that we will observe $k$ with probability $p(k|P,M)$. We aim to reproduce the statistics by using a set of states $\lambda$ that are random distributed in the set $\Lambda$ with probability $p(\lambda|P)$ every time the preparation $P$ is performed. If, for a given $\lambda$, we get the outcome $k$ with the probability $p(k|\lambda,M)$, we are able to reproduce the statistics if
\begin{equation}\label{eq con}
p(k|P,M) = \int_\Lambda p(\lambda|P) p(k|\lambda,M)d\lambda\,,
\end{equation}
and the protocol is called non-contextual if $p(\lambda|P)$ is a function of the quantum state alone, i.e., $p(\lambda|P)=p(\lambda|\rho_0)$, and $p(k|\lambda,M)$ depends only on the positive-operator-valued-measurement element $M_k$ associated to the corresponding outcome of the measurement $M$, i.e., $p(k|\lambda,M)=p(k|\lambda,M_k)$. In our case, the outcome $k$ corresponds to the work $w_k$ (which can depend on $\lambda$), and  if the protocol is non-contextual the work distribution can be expressed as
\begin{equation}\label{eq non cont}
p(w) = \sum_k p(k|P,M) \delta(w-w_k)\,,
\end{equation}
where $p(k|P,M)$ is given by Eq.~\eqref{eq con} with $p(\lambda|P)=p(\lambda|\rho_0)$ and $p(k|\lambda,M)=p(k|\lambda,M_k)$, so that for a negative quasiprobability of work we cannot have a non-contextual protocol.
For instance, the two-projective-measurement scheme gives a probability distribution that is noncontextual as observed in Ref.~\cite{lostaglio18}.
Thus, a process that cannot be reproduced within any non-contextual protocol exhibits genuinely non-classical features. We will discuss the contextuality in terms of time reversal processes in the next section, pointing to a connection with detailed fluctuation theorems.

\section{Time reversal}\label{sec. time rev}
In general, a work quasiprobability distribution $\bar{p}(w)$ for the backward process is obtained by taking in account that the initial state is a certain density matrix $\bar{\rho}$ and the events are $\bar\Pi_{i}$, $\bar\Pi_j$, $\cdots$, $\bar\Pi'_k$, $\cdots$, so that
\begin{equation}\label{eq. tr}
\bar{p}(w) = \sum_{i,j,\ldots} \bar{v}(\bar{\Pi}_i,\bar{\Pi}_j ,\ldots, \bar{\Pi}'_k,\ldots ) \delta(w-\bar{w}(\epsilon_i,\ldots))\,.
\end{equation}
In detail, the work support is given by $\bar{w}(\epsilon_i,\ldots)$ and the quasiprobability $\bar{v}(\bar{\Pi}_i,\bar{\Pi}_j ,\ldots, \bar{\Pi}'_k,\ldots )$ is calculated with respect to the initial density matrix $\bar{\rho}$, e.g., $\bar{v}(\bar{\Pi}_i,\bar{\Pi}'_k ) = \text{Re}\Tr{\bar{\Pi}_i\bar{\Pi}'_k\bar{\rho}}$ if there are only two events, which are $\bar{\Pi}_i$ and $\bar{\Pi}'_k$.
A natural choice of the backward initial state $\bar \rho$ can be the final state of the forward process, i.e., $\bar \rho = \rho'$. In this case, among all the representations of the backward process, there is always a quasiprobability distribution $\bar{p}(w)$ such that
\begin{equation}\label{eq. sym}
\bar{p}(w) = p(-w)\,,
\end{equation}
for any quasiprobability distribution $p(w)$ of the forward process.
In particular, given $p(w)$ of the form in Eq.~\eqref{eq. p}, $\bar{p}(w)$ is obtained by performing a time reversal, i.e., by replacing $\Pi_i \mapsto \bar{\Pi}_i$, $\Pi'_k \mapsto \bar{\Pi}'_k$, $\rho\mapsto \bar{\rho}$ and $w(\epsilon_i,\ldots)\mapsto \bar{w}(\epsilon_i,\ldots)$.
In detail, if $\bar \rho=\rho'$, for the quasiprobabilities we will get $\bar{v}(\bar{\Pi}_i,\bar{\Pi}_j ,\ldots, \bar{\Pi}'_k,\ldots ) =v(\Pi_i,\Pi_j ,\ldots, \Pi'_k ,\ldots) $, i.e., they are invariant under the time reversal. To prove it, it is enough to note that the quasiprobability involves the real part of a trace of the product of the projectors and the initial density matrix, e.g., $\bar{v}(\bar{\Pi}_i,\bar{\Pi}'_k ) = \text{Re}\Tr{\bar{\Pi}_i\bar{\Pi}'_k\bar{\rho}} = \text{Re}\Tr{\Pi_i \Pi'_k\rho} = v(\Pi_i,\Pi'_k ) $.
Furthermore, by requiring that the work is odd under the time reversal, we have $\bar{w}(\epsilon_i,\ldots)=-w(\epsilon_i,\ldots)$, from which we get the time-reversal symmetry relation for work in Eq.~\eqref{eq. sym}.

We note that Eq.~\eqref{eq. sym} is not satisfied for the two-projective-measurement probability distribution $p_{\text{TPM}}(w)$. Of course the probability $\Tr{\Pi_i \rho \Pi_i \Pi'_k}$ in Eq.~\eqref{eq. TPM} is of the form $v_{TPM}(E,F)=\Tr{E F E \rho}$, so that does not satisfy the Gleason axiom in Eq.~\eqref{Q3}.
Let us focus on the forward class of quasiprobability distributions in Eq.~\eqref{eq. pq}, which reproduce the two-projective-measurement scheme when the initial state $\rho$ is incoherent with respect to the projectors $\Pi_i$, i.e., $p_q(w)=p_{\text{TPM}}(w)$ when $\rho=\Delta(\rho)$ for any $q$.
Similarly, the backward class satisfying the conditions (W1),(W2) and (W3) is formed by the quasiprobability distributions
\begin{equation}\label{eq. pq tilde}
\tilde{p}_q(w) = \sum_{i,k,l} \text{Re}\Tr{\bar{\Pi}'_k \bar{\rho} \bar{\Pi}'_l \bar{\Pi}_i} \delta(w- \epsilon_i+q \epsilon'_k + (1-q)\epsilon'_l)\,,
\end{equation}
with $q$ real, such that $\tilde{p}_q(w)=\bar{p}_{\text{TPM}}(w)$ when $\bar{\rho}=\bar{\Delta}(\bar{\rho})$ for any $q$.
On the other hand, the time reversal of the quasiprobability distribution $p_q(w)$ such that Eq.~\eqref{eq. sym}  holds, reads
\begin{equation}\label{eq. pq bar}
\bar{p}_q(w) = \sum_{i,j,k} \text{Re}\Tr{\bar{\Pi}_i \bar{\rho} \bar{\Pi}_j \bar{\Pi}'_k} \delta(w- q \epsilon_i - (1-q)\epsilon_j+\epsilon'_k)\,,
\end{equation}
which reproduces the two-projective-measurement scheme, $\bar{p}_q(w)=\bar{p}_{\text{TPM}}(w)$ when $\bar{\rho}=\bar{\Delta}(\bar{\rho})$, for $q=0,1$ but not for all $q$ (in particular, in Ref.~\cite{Pei23} it has been originally observed how $p_q(w)$ does not satisfy the symmetry relation in Eq.~\eqref{eq. sym} for $q\neq0,1$, if we use $\tilde p_q(w)$ instead of $\bar p_q(w)$ in this relation).
%, with $\bar{\Delta}(\bar{\rho}) = \sum_k \bar{\Pi}'_k\bar{\rho}\bar{\Pi}'_k$.
%We note that $\bar{p}_{\text{TPM}}(w)$ is not obtained from $p_{\text{TPM}}(w)$, unlike what was done for the quasiprobability distribution of work, since  the two-projective measurements scheme  is defined from an operational procedure. However for incoherent initial states $\rho=\Delta(\rho)$  and $\bar{\rho}=\bar{\Delta}(\bar{\rho})$, we get the symmetry relation $\bar{p}_{\text{TPM}}(w)=p_{\text{TPM}}(-w)$ since the two probability distributions are related by a time reversal transformation $\Pi_i \mapsto \bar{\Pi}_i$, $\Pi'_k \mapsto \bar{\Pi}'_k$, $\rho\mapsto \bar{\rho}$ and $\epsilon'_k-\epsilon_i\mapsto \epsilon_i-\epsilon'_k$.
%Thus, since in general $\bar{p}_q(w)\neq\bar{p}_{\text{TPM}}(w)$ when $\bar{\rho}=\bar{\Delta}(\bar{\rho})$,
Thus, the forward class with the quasiprobability distributions of Eq.~\eqref{eq. pq} is not mapped into the  backward class with the quasiprobability distributions of Eq.~\eqref{eq. pq tilde} by performing the time reversal. In detail,  we get $\tilde{p}_q(w) = \bar{p}_q(w) = p_q(-w)$ for $q=0,1$, but in general $\tilde{p}_q(w) \neq\bar{p}_q(w)= p_q(-w)$ for $q\neq 0,1$.
The symmetry relation in Eq.~\eqref{eq. sym} relates $p_q(w)$ with $\bar{p}_q(w)$, and not with $\tilde p_q(w)$. Then, the negativity of $\tilde p_q(w)$ is not constrained by the negativity of $p_q(w)$ for $q\neq 0,1$.
This result means that the (non)negativity of the forward class does not imply the (non)negativity of the backward class, and vice versa.
The negativity is related to genuine quantum features: If there is some nonnegative distribution in the forward (backward) class, the forward (backward) work statistics can be reproduced with a noncontextual hidden variable model%~\cite{Spekkens05,Spekkens15,lostaglio18} 
 which satisfies the conditions (W1), (W2) and (W3).
Let us show it by giving an example of a forward protocol that is noncontextual but it shows contextuality in its time reversal.
We consider a one dimensional system, with initial Hamiltonian $H(0)= x$ and final Hamiltonian $H(\tau)= p^2$, where $x$ is the position and $p$ is the momentum, such that $[x,p]=i$. In this case we consider the projectors $\Pi_x=\ket{x}\bra{x}$ and $\Pi'_p=\ket{p}\bra{p}$ and the sudden time evolution $U_{0,\tau}=I$. It is easy to show that for $q=1/2$ the  quasiprobability distribution of work can be expressed in terms of the Wigner function $W(x,p)$ as (see Ref.~\cite{Francica222})
\begin{equation}\label{eq 1/2}
p_{1/2}(w) = \int dx dp W(x,p) \delta(w- p^2+ x)\,.
\end{equation}
We consider the initial wave function $\braket{x}{\psi}=\exp(-ax^2+bx+c)$, we get $W(x,p)\geq 0$, and the protocol is noncontextual.
However, for this state we get the backward quasiprobability distribution
\begin{equation}
\tilde p_{q}(w) = \int dp dp' \tilde v(p,w+qp^2+(1-q){p'}^2,p')\,,
\end{equation}
where $\tilde v(p,x,p')=\text{Re} \braket{p}{\psi}\braket{\psi}{p'}\braket{p'}{x}\braket{x}{p}$. Then, $\tilde p_{q}(w)$ takes also negative values for any $q$, so that $\nexists q$ such that $\tilde p_q(w)\geq 0$ for all $w$, and there is contextuality for the backward process.
Finally, we will aim to generalize the detailed fluctuation theorem of Eq.~\eqref{eq. FT crooks} to quasiprobabilities in the next section. Then, for a given $p_q(w)$ we will consider the time-reversed $\bar p_q(w)$ with the same support of $p_q(-w)$ and an appropriate initial state $\bar \rho$ for the backward process.
%Although $\bar p_q(w)$ satisfies the condition (W1) only for $q=0,1$, a detailed fluctuation relation of the form of Eq.~\eqref{eq. FT crooks} will relate $p_q(w)$ and $\bar p_q(-w)$, since they have the same support and not $p_q(w)$ and $\tilde p_q(-w)$ for $q\neq 0,1$.
To do this, we must also take into account quantum coherence as a random variable as done in Ref.~\cite{Francica22}.

\section{Detailed Fluctuation Theorem with initial quantum coherence}\label{sec. deta}
Given a dephasing map $\Delta$, the quantum coherence of a state $\rho$ can be characterized by using the relative entropy of coherence (see, e.g., the review in Ref.~\cite{Streltsov17})
\begin{equation}
C_\Delta(\rho) = S(\Delta(\rho))-S(\rho)\,,
\end{equation}
where we have introduced the von Neumann entropy $S(\rho) = - \Tr{\rho \ln \rho}$. Let us focus on the forward process. By considering the eigenvalues $r_n$ and the eigenstates $\ket{r_n}$ of the initial state $\rho$, such that $\rho = \sum r_n R_n$, where $R_n=\ket{r_n}\bra{r_n}$, we define the probability distribution of coherence~\cite{Francica22}
\begin{equation}
p_c(C) = \sum_{i,n} r_n \Tr{R_n \Pi_i}\delta(C+\ln\bra{\epsilon_i}\rho\ket{\epsilon_i}-\ln r_n)\,,
\end{equation}
such that $C_\Delta (\rho)=\langle C \rangle=\int C p_c(C) dC$. We note that $r_n \Tr{R_n \Pi_i}=\Tr{\rho R_n \Pi_i} = v(R_n,\Pi_i)$, which is nonnegative since $[\rho,R_n]=0$ for all $n$. Thus, the state is $\rho$ and the events are $R_n$ and $\Pi_i$.
In the presence of initial quantum coherence, the work  can be represented by the quasiprobability distribution $p_q(w)$.
To derive a detailed fluctuation theorem, we consider an initial state $\rho$ such that its incoherent part (with respect to the energy basis) is thermal, $\Delta(\rho)=\rho_\beta$. In this case, we get the integral fluctuation relation of Ref.~\cite{Francica22},
\begin{equation}\label{eq. FT integral}
\langle e^{-\beta(w-\Delta F) - C}\rangle = 1\,,
\end{equation}
which is our starting point to derive the detailed fluctuation theorem. In detail, the average in Eq.~\eqref{eq. FT integral} is calculated with respect  to the joint quasiprobability distribution
\begin{eqnarray}\label{eq. q qp}
\nonumber && p_{q,q'}(w,C) = \sum_{k,j,i,n} r_n \text{Re} \Tr{R_n \Pi_j \Pi'_k \Pi_i} \delta(w-\epsilon'_k+q\epsilon_i\\
\nonumber && +(1-q)\epsilon_j)\delta(C+q'\ln\bra{\epsilon_i}\rho\ket{\epsilon_i}+(1-q')\ln\bra{\epsilon_j}\rho\ket{\epsilon_j}\\
 &&  -\ln r_n)\,.
\end{eqnarray}
We can easily check that the marginal distributions are the quasiprobability distribution of work $p_q(w)=\int  p_{q,q'}(w,C) dC$ and the probability distribution of initial quantum coherence $p_c(C)=\int  p_{q,q'}(w,C) dw$.
To formulate a detailed fluctuation theorem, we focus on the quantity $e^{-\beta(w-\Delta F) - C} p_{q,q'}(w,C)$.
Only for $q=q'$, by considering $ p_{q}(w,C)= p_{q,q}(w,C)$, we get that $e^{-\beta(w-\Delta F) - C} p_{q}(w,C)$ is a joint quasiprobability distribution,
\begin{equation}\label{eq. FT prototype}
e^{-\beta(w-\Delta F) - C} p_{q}(w,C)=\hat p_{q}(-w,C)\,,
\end{equation}
which explicitly reads
\begin{eqnarray}\label{eq. p hat}
\nonumber && \hat p_{q}(w,\hat C) =  \sum_{k,j,i,n} \frac{e^{-\beta \epsilon'_k}}{Z'} \text{Re} \Tr{\bar\Pi'_k \bar\Pi_i\bar R_n \bar\Pi_j }\delta(w-q\epsilon_i \\
\nonumber && -(1-q)\epsilon_j+\epsilon'_k)\delta(\hat C+q\ln\bra{\epsilon_i}\rho\ket{\epsilon_i} +(1-q) \ln\bra{\epsilon_j}\rho\ket{\epsilon_j}\\
&&-\ln r_n)\,.
\end{eqnarray}
The quasiprobability distribution $\hat p_{q}(w,\hat C)$ represents a backward process, where the initial state is $\bar\rho=\rho'_\beta$ and the events are $\bar \Pi'_k$,$\bar\Pi_i$, $\bar R_n=U_{\tau,0} R_n U_{\tau,0}^\dagger$ and $\bar\Pi_j$. In particular, the projector $\bar \Pi'_k$ selects the pure state $\bar\Pi'_k$ with  probability $e^{-\beta \epsilon'_k}/Z'$.
We note that one of the marginal distribution is equal to the two-projective-measurement probability distribution for the backward process starting from $\bar \rho$ such that $\bar\Delta(\bar\rho)=\rho'_\beta$, $\int \hat p_{q}(w,\hat C)d\hat C=\bar{p}_{TPM}(w)$, so that by integrating Eq.~\eqref{eq. FT prototype} over $C$, and noting that in general $\int e^{- C} p_{q,q'}(w,C)dC=p_{TPM}(w)$, we get the Tasaki-Crooks fluctuation relation in Eq.~\eqref{eq. FT crooks}. Concerning the variable $\hat C$ of the backward process, its average is
\begin{equation}
\int \hat C \hat p_{q}(\hat C) d\hat C = S( U^\dagger_{0,\tau}\rho'_\beta U_{0,\tau}|| \rho_\beta) - S( U^\dagger_{0,\tau} \rho'_\beta U_{0,\tau}||\rho)\,,
\end{equation}
where $\hat p_{q}(\hat C)=\int \hat p_{q}(w,\hat C) dw$ and the quantum relative entropy is defined as $S(\rho||\eta)=-S(\rho)-\Tr{\rho \ln \eta}$.

We aim to get a more symmetric detailed fluctuation relation, which implies Eq.~\eqref{eq. FT prototype}.
We are looking for a fluctuation relation of the form $e^{a\cdot x} p(x)=\bar p(\bar{x})$, where $x$ is a set of variables including the work $w$, e.g., $x=(w,C,\ldots)$, $p(x)$ is the forward distribution and $\bar p(\bar{x})$ is the backward distribution of $\bar{x}=(-w,\bar{C},\ldots)$. It is worth noting that the positive multiplying factor $e^{a\cdot x}$ does not change the support of $p(x)$, then the backward distribution $\bar p(\bar{x})$ has the same support of $p(x)$. This suggests that the time reversal discussed in the previous section, i.e., $\bar p_q(w)$ in Eq.~\eqref{eq. pq bar} with an appropriate initial state $\bar \rho$, will play some role.
For simplicity, we consider $q=0$, so that, from Eq.~\eqref{eq. q qp}, we get
\begin{eqnarray}
\nonumber && p_{0}(w,C) = \sum_{k,i,n} r_n \text{Re} \Tr{R_n \Pi_i \Pi'_k } \delta(w-\epsilon'_k+\epsilon_i)\\
 && \times\delta(C+\ln\bra{\epsilon_i}\rho\ket{\epsilon_i} -\ln r_n)\,.
\end{eqnarray}
Since the backward distribution in the right side of a detailed fluctuation relation needs to have the same support of the left side, we try to introduce a random variable $\bar C$, and the joint time-reversed quasiprobability distribution of the form
\begin{eqnarray}
\nonumber && \bar{p}_{0}(w,\bar C) = \sum_{k,i,n} \bar{r}_n \text{Re} \Tr{\bar R_n \bar \Pi_i \bar \Pi'_k } \delta(w-\epsilon_i+\epsilon'_k)\\
 && \times\delta(\bar{C}+\ln\bra{\epsilon_i}\rho\ket{\epsilon_i} -\ln r_n)\,,
\end{eqnarray}
such that $\bar{p}_{0}(-w,C)$ has the same support of $p_{0}(w,C)$.
The eigenvalues $\bar r_n$ and the projectors $\bar R_n$ are such that $\bar{\rho}=\sum \bar r_n \bar R_n$, where the initial state $\bar \rho$ of the backward process can be chosen appropriately in order to get a detailed fluctuation relation. E.g., for $\rho=\rho_\beta$ and $\bar \rho = \rho'_\beta$ we get the detailed fluctuation theorem of Eq.~\eqref{eq. FT crooks}. However, as just seen in Eq.~\eqref{eq. FT prototype},
\begin{equation}\label{eq. ineq}
e^{-\beta(w-\Delta F) - C}p_{0}(w,C)\neq \bar{p}_{0}(-w,C)
\end{equation}
for any $\bar \rho$ and the two quasiprobability distributions are not related by a detailed fluctuation theorem.
Of course, the equality in Eq.~\eqref{eq. ineq} is achieved when the initial state is incoherent, i.e., $\rho=\Delta(\rho)$, so that $\bar{p}_{0}(w,\bar{C})=\hat{p}_{0}(w,\bar{C})=\bar{p}_{TPM}(w) \delta(\bar C)$. To get a fluctuation theorem we start to focus on the variable $\bar{C}$, which for the backward process has the marginal probability distribution $\bar p(\bar C)=\int\bar{p}_{0}(w,\bar C)dw$, which reads
\begin{equation}
\bar{p}(\bar C) = \sum_{i,n} \bar{r}_n  \Tr{\bar R_n \bar \Pi_i }\delta(\bar{C}+\ln\bra{\epsilon_i}\rho\ket{\epsilon_i} -\ln r_n)\,.
\end{equation}
We guess that for the forward process, the variable $\bar C$ has the probability distribution
\begin{equation}\label{eq. bar C prob}
p(\bar C) = \sum_{k,n} r_n  \Tr{ R_n  \Pi'_k }\delta(\bar{C}+\ln \bra{\epsilon'_k}\bar{\rho}\ket{\epsilon'_k}-\ln \bar{r}_n)\,.
\end{equation}
Thus, by introducing this new random variable, we consider the joint quasiprobability distribution
\begin{eqnarray}\label{eq. p0 F}
\nonumber && p_{0}(w,C,\bar C) = \sum_{k,i,n} r_n \text{Re} \Tr{R_n \Pi_i \Pi'_k } \delta(w-\epsilon'_k+\epsilon_i)\\
\nonumber && \times\delta(C+\ln\bra{\epsilon_i}\rho\ket{\epsilon_i} -\ln r_n)\delta(\bar{C}+\ln \bra{\epsilon'_k}\bar{\rho}\ket{\epsilon'_k}\\
 &&-\ln \bar{r}_n)
\end{eqnarray}
and the time-reversed one
\begin{eqnarray}\label{eq. p0 B}
\nonumber && \bar{p}_{0}(w,C,\bar C) = \sum_{k,i,n} \bar{r}_n \text{Re} \Tr{\bar R_n \bar \Pi_i \bar \Pi'_k } \delta(w-\epsilon_i+\epsilon'_k)\\
\nonumber && \times\delta(C+\ln \bra{\epsilon'_k}\bar{\rho}\ket{\epsilon'_k}-\ln \bar{r}_n)\delta(\bar{C}+\ln\bra{\epsilon_i}\rho\ket{\epsilon_i}\\
  &&-\ln r_n)\,,
\end{eqnarray}
so that $p_{0}(w,C,\bar C)$ and $\bar{p}_{0}(-w,\bar C,C)$ have the same support. It is easy to generalize Eqs.~\eqref{eq. p0 F}-\eqref{eq. p0 B} for $q\neq 0$, and to check that these two quasiprobability distributions are related by the detailed fluctuation relation
\begin{equation}\label{eq. FT q}
e^{-\beta(w-\Delta F) - C+\bar C}p_{q}(w,C,\bar C)= \bar{p}_{q}(-w,\bar C, C)\,,
\end{equation}
if $\Delta(\rho)=\rho_\beta$, $\bar \Delta(\bar\rho)=\rho'_\beta$ and there is also the relation $\bar{R}_n = U_{0,\tau}R_{\pi_n}U_{0,\tau}^\dagger$ for all $n$ with some permutation $\pi_n$ of the indices (i.e., $\rho'$ and $\bar{\rho}$ are diagonal with respect to the same basis). We note that
\begin{equation}\label{eq. rel C hat C bar}
\int e^{-C} \bar{p}_{q}(w,C, \bar C) d C = \hat p_q(w,\bar C)\,,
\end{equation}
then  the fluctuation theorem of Eq.~\eqref{eq. FT q} implies the relation in Eq.~\eqref{eq. FT prototype} (it is enough to multiply Eq.~\eqref{eq. FT q} by $e^{-\bar C} $  and then to integrate over $\bar C$). %In particular, Eq.~\eqref{eq. rel C hat C bar} defines $\hat C$ in terms of $\bar C$.
Given the initial state $\rho$, there are several states $\bar{\rho}$ such that $\bar{\Delta}(\bar{\rho})=\rho'_\beta$, but in general there can be a unique  $\bar{\rho}$  if we require also that $\bar{R}_n = U_{0,\tau}R_{\pi_n}U_{0,\tau}^\dagger$, which is $\bar{\rho} = \sum \bar{r}_n U_{0,\tau}R_{\pi_n}U_{0,\tau}^\dagger$ where the eigenvalues $\bar{r}_n$ are the solutions of the linear equations
\begin{equation}\label{eq. rbn}
\sum_n \bar{r}_n \bra{\epsilon'_k}U_{0,\tau}R_{\pi_n}U_{0,\tau}^\dagger\ket{\epsilon'_k} = e^{-\beta \epsilon'_k}/Z'
\end{equation}
with $k=1,2,\ldots$. While the condition $\bar{\Delta}(\bar{\rho})=\rho'_\beta$ determines the populations of $\bar{\rho}$, the condition $U^\dagger \bar{R}_n U = R_{\pi_n}$ for all $n$ fixes the coherences of $\bar{\rho}$.  However, these two conditions are not always compatible, i.e., Eq.~\eqref{eq. rbn} can also have solutions $\bar{r}_n$ which are negative, $\bar {r}_{n}<0$, for some $n$. In this case such initial state $\bar \rho$ does not exist, since $\bar \rho \geq 0$.
Furthermore, we note that for an incoherent state $\rho=\Delta(\rho)$ there is coherence in the state $\bar{\rho}$, namely $\bar{\Delta}(\bar{\rho})\neq \bar{\rho}$, thus for $\rho=\rho_\beta$ the backward process starts from the initial state $\bar \rho\neq \rho'_\beta$ and it is different from the Tasaki-Crooks backward process of Sec.~\ref{sec. incohe}. However, if $\bar{R}_n = U_{0,\tau}R_{\pi_n}U_{0,\tau}^\dagger$ is satisfied together with $\Delta(\rho)=\rho_\beta$ and $\bar \Delta(\bar\rho)=\rho'_\beta$, Eq.~\eqref{eq. FT q} implies Eq.~\eqref{eq. FT prototype} for the trivial permutation $\pi_n=n$, which implies the Tasaki-Crooks fluctuation relation in Eq.~\eqref{eq. FT crooks}.
In general, for $\bar{R}_n \neq U_{0,\tau}R_{\pi_n}U_{0,\tau}^\dagger$, the detailed fluctuation relation in Eq.~\eqref{eq. FT q} holds if $\Delta(\rho)=\rho_\beta$ and $\bar \Delta(\bar\rho)=\rho'_\beta$, for
\begin{eqnarray}\label{eq. p F-q}
\nonumber && p_{q}(w,C,\bar C) = \sum_{k,i,j,n,m} r_n \text{Re} \Tr{R_n U_{0,\tau}^\dagger \bar R_m U_{0,\tau} \Pi_i \Pi'_k \Pi_j } \\
\nonumber &&\times \delta(w-\epsilon'_k+q\epsilon_i+(1-q)\epsilon_j)\delta(C+q\ln\bra{\epsilon_i}\rho\ket{\epsilon_i}+(1-q)\\
 && \times \ln\bra{\epsilon_j}\rho\ket{\epsilon_j}-\ln r_n)\delta(\bar{C}+\ln \bra{\epsilon'_k}\bar{\rho}\ket{\epsilon'_k}-\ln \bar{r}_m)
\end{eqnarray}
and
\begin{eqnarray}\label{eq. p B-q}
\nonumber && \bar{p}_{q}(w,C,\bar C) = \sum_{k,i,j,n,m} \bar{r}_m \text{Re} \Tr{U_{0,\tau} R_nU_{0,\tau}^\dagger \bar R_m \bar \Pi_i \bar \Pi'_k \bar \Pi_j } \\
\nonumber &&\times \delta(w-q\epsilon_i-(1-q)\epsilon_j+\epsilon'_k) \delta(C+\ln \bra{\epsilon'_k}\bar{\rho}\ket{\epsilon'_k}-\ln \bar{r}_m)\\
 && \times \delta(\bar{C}+q\ln\bra{\epsilon_i}\rho\ket{\epsilon_i}+(1-q)\ln\bra{\epsilon_j}\rho\ket{\epsilon_j}-\ln r_n)\,,
\end{eqnarray}
which are linear in the respective initial states.
%which reduce to Eqs.~\eqref{eq. p0 F} and~\eqref{eq. p0 B} for $q=0$ if $\bar{\rho}$ can be chosen such that $\bar\rho$ and $\rho'$ are diagonal in the same basis (which is not always possible), i.e.,  $R_n U_{0,\tau}^\dagger \bar R_m U_{0,\tau}=\delta_{n,\pi_m}R_n$ for some permutation $\pi_n$.
Thus, we note that for $\rho=\rho_\beta$ we can consider a backward process starting from the initial state $\bar \rho= \rho'_\beta$ and in this case Eq.~\eqref{eq. FT q} reduces to Eq.~\eqref{eq. FT crooks} for $q=0$. In general, Eq.~\eqref{eq. rel C hat C bar} still holds, thus Eq.~\eqref{eq. FT q} implies Eq.~\eqref{eq. FT prototype}.
We note that although Eq.~\eqref{eq. FT q} defines a relation between quasiprobability distributions, it suggests that for values of $w$, $C$ and $\bar C$ such that $\beta(w-\Delta F) + C-\bar C>0$, the time-reversal quasiprobability distribution $\bar{p}_{q}(-w,\bar C, C)$ is exponentially suppressed and it is practically zero. Thus, these values of $w$, $C$ and $\bar C$ are practically absent in the statistics of the time-reversed process represented by $\bar{p}_{q}(-w,\bar C, C)$, although they are present for the forward process. This helps to clarify how a thermodynamic time arrow can still emerge from quasiprobabilities in the presence of initial coherence, and it is affected by the latter.
Furthermore, from Eq.~\eqref{eq. p F-q} we get a marginal quasiprobability distribution $p(\bar C) = \int p_q(w,C,\bar C)dwdC$, which explicitly reads
\begin{equation}\label{eq. bar C prob quasi}
p(\bar C) = \sum_{k,m} \text{Re} \Tr{\rho' \bar R_m \bar\Pi'_k }\delta(\bar{C}+\ln \bra{\epsilon'_k}\bar{\rho}\ket{\epsilon'_k}-\ln \bar{r}_m)
\end{equation}
and reduces to the probability distribution in Eq.~\eqref{eq. bar C prob} if the condition $\bar{R}_n = U_{0,\tau}R_{\pi_n}U_{0,\tau}^\dagger$ is satisfied. In particular, in the next section we will show that in this case a bound for the average work is intimately related to the final quantum coherence.

\section{Integral fluctuation theorems and bounds}\label{sec. inte}
From the detailed fluctuation theorem of Eq.~\eqref{eq. FT q} we can derive two integral fluctuation relations. By multiplying  Eq.~\eqref{eq. FT q} by $e^{-\bar{C}}$ and integrating we get Eq.~\eqref{eq. FT integral}. Furthermore, by integrating Eq.~\eqref{eq. FT q}, we get
\begin{equation}\label{eq. int ft 2}
\langle e^{-\beta(w-\Delta F)-C + \bar{C}}\rangle = 1\,.
\end{equation}
The integral fluctuation theorem of Eq.~\eqref{eq. FT integral} has been formulated in Ref.~\cite{Francica22} and further discussed in Ref.~\cite{Francica23}. In particular, it implies the bound
\begin{equation}\label{eq. bound 1}
\beta(\langle w\rangle-\Delta F)+ \langle C\rangle \geq 0\,.
\end{equation}
Concerning Eq.~\eqref{eq. int ft 2},  we note that $\beta(w-\Delta F)+C$ can be replaced with a random variable $\sigma$, i.e., for any function of two variables $f(x,y)$,
\begin{eqnarray}
\nonumber &&\langle f(\beta(w-\Delta F)+C,\bar{C})\rangle=\int f(\beta(w-\Delta F)+C,\bar{C})\\
\nonumber && \times p_q(w,C,\bar{C})d w d C d\bar{C}=\int f(\sigma,\bar{C}) p(\sigma,\bar{C}) d\sigma d \bar{C}  \\
&&=  \langle f(\sigma,\bar{C})\rangle\,,
\end{eqnarray}
where
\begin{eqnarray}
\nonumber p(\sigma, \bar{C})&=&  \sum_{n,m,k} r_n \Tr{\Pi'_k R_n U_{0,\tau}^\dagger \bar R_m U_{0,\tau}} \delta(\sigma-\beta \epsilon'_k -\ln Z'\\
&&-\ln r_n)  \delta(\bar{C}+\ln \bra{\epsilon'_{k}}\bar{\rho}\ket{\epsilon'_{k}}-\ln \bar{r}_m)\,.
\end{eqnarray}
In particular, if $\bar{R}_n = U_{0,\tau}R_nU_{0,\tau}^\dagger$, we get the probability distribution
\begin{eqnarray}
\nonumber p(\sigma, \bar{C})&=&  \sum_{n,k} r_n \Tr{\Pi'_k R_n} \delta(\sigma-\beta \epsilon'_k -\ln Z'-\ln r_n) \delta(\bar{C}\\
&&+\ln \bra{\epsilon'_{k}}\bar{\rho}\ket{\epsilon'_{k}}-\ln \bar{r}_n)\,.
\end{eqnarray}
Thus, in this case the fluctuation relation in Eq.~\eqref{eq. int ft 2} is equivalent to $\langle e^{-\sigma + \bar{C}}\rangle = 1$ and by using the Jensen's theorem we get $\langle \sigma\rangle -\langle \bar C \rangle\geq 0$, which can be expressed as
\begin{equation}\label{eq. bound 2}
\beta(\langle w\rangle-\Delta F)+ \langle C\rangle- \langle \bar{C}\rangle \geq 0\,,
\end{equation}
since $\langle \sigma \rangle = \beta(\langle w\rangle-\Delta F)+ \langle C\rangle$. Eq.~\eqref{eq. bound 2} holds even if $\bar{R}_n\neq U_{0,\tau}R_{\pi_n}U_{0,\tau}^\dagger$ and $p(\sigma, \bar{C})$ takes also negative values, since in this case $\sigma-\bar C$ can be replaced with a random variable $\bar \sigma$ having the probability distribution
\begin{equation}
p(\bar \sigma) = \sum_{n,m}r_n \Tr{U_{\tau,0}R_nU_{\tau,0}^\dagger \bar R_m}\delta(\bar \sigma - \ln r_n + \ln \bar r_m)\,,
\end{equation}
so that we obtain $\langle e^{-\bar \sigma}\rangle = 1$ and, by using the Jensen's theorem, $\langle \bar\sigma \rangle \geq 0$, i.e., Eq.~\eqref{eq. bound 2}.
Eqs.~\eqref{eq. bound 1} and~\eqref{eq. bound 2} give two lower bounds for the average work $\langle w\rangle$. Which of the two is tighter depends on the sign of $\langle \bar{C} \rangle$.
By noting that
\begin{equation}\label{eq. b2}
\langle \sigma \rangle -\langle \bar{C} \rangle = S(\rho'|| \bar{\rho})\geq 0\,,\quad \langle \sigma \rangle = S(\rho'|| \rho'_\beta)\geq 0\,,
\end{equation}
$\langle \bar{C} \rangle$  can be expressed as the difference of quantum relative entropies
\begin{equation}\label{eq. Cbar}
\langle \bar{C}\rangle = S(\rho'|| \rho'_\beta)-S(\rho'|| \bar{\rho})\,.
\end{equation}
The same result is also achieved by averaging with respect to the quasiprobability in Eq.~\eqref{eq. bar C prob quasi}.
In particular, if the eigenvalues $r_n$ and $\bar r_m$ are sorted in decreasing order, i.e., $r_n\geq r_{n+1}$ and $\bar r_m \geq \bar r_{m+1} $, from the von Neumann's trace inequality we get $\Tr{\rho' \ln \bar \rho} \leq \sum r_n \ln \bar r_n$, from which $S(\rho'|| \bar{\rho})\geq \sum r_n \ln (r_n/\bar r_n)$, where the equality holds if $\bar{R}_n=U_{0,\tau}R_n U_{0,\tau}^\dagger$, so that $\langle \bar{C}\rangle $ is maximum and the bound of Eq.~\eqref{eq. bound 2} is tighter in this case.
We note that $\langle \bar{C} \rangle$ is related to the final quantum coherence in the energy basis, quantified by $C_{\bar{\Delta}}(\rho')= S(\bar{\Delta}(\rho'))-S(\rho')$. In particular, Eq.~\eqref{eq. Cbar} can be written as
\begin{equation}\label{eq. bar C 2}
\langle \bar{C}\rangle = C_{\bar{\Delta}}(\rho') + S(\bar{\Delta}(\rho')||\bar{\Delta}(\bar{\rho})) -S(\rho'||\bar{\rho})\,.
\end{equation}
To prove it, it is enough to note that~\cite{Francica19}
\begin{eqnarray}
\nonumber && S(\rho'||\rho'_\beta) = S(\rho'||\bar{\Delta}(\bar{\rho})) = -S(\rho') - \Tr{\rho' \ln \bar{\Delta}(\bar{\rho})}\\
\nonumber &&=  S(\bar{\Delta}(\rho'))-S(\rho') -S(\bar{\Delta}(\rho'))-  \Tr{\bar{\Delta}(\rho') \ln \bar{\Delta}(\bar{\rho})} \\
 &&= C_{\bar{\Delta}}(\rho') + S(\bar{\Delta}(\rho')||\bar{\Delta}(\bar{\rho}))\,.
\end{eqnarray}
From Eq.~\eqref{eq. Cbar}, we get the bounds for $\langle \bar C \rangle$
\begin{equation}\label{eq. bounds 0}
-S(\rho'|| \bar{\rho})\leq\langle \bar{C}\rangle \leq S(\rho'|| \rho'_\beta)\,,
\end{equation}
whereas, from Eq.~\eqref{eq. bar C 2}, we get
\begin{equation}\label{eq. bounds}
S(\bar{\Delta}(\rho')||\bar{\Delta}(\bar{\rho}))-S(\rho'||\bar{\rho})\leq\langle \bar{C}\rangle \leq C_{\bar{\Delta}}(\rho')\,,
\end{equation}
since $\bar{\Delta}$ is a completely positive and trace preserving map, so that $S(\bar{\Delta}(\rho')||\bar{\Delta}(\bar{\rho}))-S(\rho'||\bar{\rho})\leq 0$ and $ C_{\bar{\Delta}}(\rho')\geq 0$.
%From Eq.~\eqref{eq. bar C 2}, we note that Eq.~\eqref{eq. bound 2} can be expressed as
%\begin{equation}
%\beta(\langle w \rangle - \Delta F ) \geq C_{\bar{\Delta}}(\rho') - C_\Delta(\rho) + S(\bar{\Delta}(\rho')||\bar{\Delta}(\bar{\rho}))- S(\rho'||\bar{\rho})
%\end{equation}
The bounds in Eq.~\eqref{eq. bounds 0} and in Eq.~\eqref{eq. bounds} can be saturated, depending on the final state $\rho'$. If $\rho'=\bar \rho$, i.e., $r_{\pi_n}=\bar r_n$ and $\bar{R}_n= U_{0,\tau}R_{\pi_n}U_{0,\tau}^\dagger$, then $\langle \bar C\rangle = S(\rho'|| \rho'_\beta)= C_{\bar \Delta}(\rho')$. In this case, $p(\bar C)$ reduces to the probability distribution of coherence of $\rho'$ with respect to the final energy basis and $\bar p(\bar C) = p_c(\bar C)$. In contrast, the lower bound in Eq.~\eqref{eq. bounds} can be saturated only if $C_{\bar \Delta}(\rho')=0$, i.e., $\rho'=\bar\Delta(\rho')$. Let us focus on $\bar{R}_n= U_{0,\tau}R_{\pi_n}U_{0,\tau}^\dagger$. In this case the bound is zero since, if $\rho'=\bar\Delta(\rho')$, from Eq.~\eqref{eq. rbn} we get $\bar{\rho}=\bar\Delta(\bar \rho)$, from which $S(\rho'||\bar{\rho})=S(\bar{\Delta}(\rho')||\bar{\Delta}(\bar{\rho}))$. Instead, the lower bound of Eq.~\eqref{eq. bounds 0} is saturated if $\rho'=\rho'_\beta$, then $\langle \bar C\rangle=-S(\rho'|| \bar{\rho})=0$ since $\rho'=\bar\Delta(\rho')$ and thus $\rho'=\bar\rho$.
From Eq.~\eqref{eq. bounds}, by noting that both the lower and upper bounds are zero if $\rho'=\bar{\Delta}(\rho')$ and $\bar{R}_n= U_{0,\tau}R_{\pi_n}U_{0,\tau}^\dagger$, then, if $\bar{R}_n= U_{0,\tau}R_{\pi_n}U_{0,\tau}^\dagger$, we get
\begin{equation}
\rho'=\bar{\Delta}(\rho') \Rightarrow \langle \bar{C} \rangle = 0\,.
\end{equation}
However, $\langle \bar{C} \rangle = 0 \nRightarrow \rho'=\bar{\Delta}(\rho')$.
\begin{proof}
To show it, we note that
\begin{eqnarray}
\nonumber \langle \bar{C}\rangle  &=& \Tr{\rho' \ln \bar{\rho}}- \Tr{\bar{\Delta}(\rho')\ln \bar{\Delta}(\bar{\rho})}\\
  &=& \sum_n r_{\pi_n}\ln\bar{r}_n - \sum_k \bra{\epsilon'_k} \rho'\ket{\epsilon'_k} \ln p'_{eq,k}\,,
\end{eqnarray}
where $p'_{eq,k}=e^{-\beta \epsilon'_k}/Z'$. Let us consider $U_{\tau,0}$ such that $U_{\tau,0}\ket{r_1}= \sqrt{a} \ket{\epsilon'_1}+\sqrt{1-a}\ket{\epsilon'_2}$, $U_{\tau,0}\ket{r_2}= \sqrt{a} \ket{\epsilon'_2}-\sqrt{1-a}\ket{\epsilon'_1}$ and $U_{\tau,0}\ket{r_n}=\ket{\epsilon'_n}$ for $n>2$. We focus on $a=1-\eta$ and $\eta\to 0$. From Eq.~\eqref{eq. rbn}, if $\pi_n=n$ we get the solution $\bar{r}_1 = p'_{eq,1}+\eta$ and $\bar{r}_2 = p'_{eq,2}-\eta$. Furthermore, $\bra{\epsilon'_1} \rho'\ket{\epsilon'_1}=r_1+(r_2-r_1)\eta$ and $\bra{\epsilon'_2} \rho'\ket{\epsilon'_2}=r_2-(r_2-r_1)\eta$, then
\begin{equation}\label{eq. ceta}
\langle \bar{C}\rangle  \simeq \left( \frac{r_1}{p'_{eq,1}}- \frac{r_2}{p'_{eq,2}}+(r_1-r_2)\ln\left(\frac{p'_{eq,1}}{p'_{eq,2}}\right)\right) \eta\,,
\end{equation}
which can be negative or positive depending on the eigenvalues $r_n$ and the populations $p'_{eq,k}$. It is easy to see that $\langle \bar{C}\rangle  \simeq0 $ if
$r_1 = P\equiv (p'_{eq,1}+p'_{eq,1} p'_{eq,2} \ln (p'_{eq,1}/p'_{eq,2}))/(1+2 p'_{eq,1} p'_{eq,2} \ln (p'_{eq,1}/p'_{eq,2}))$, although $\eta>0$ and so $\rho'\neq\bar{\Delta}(\rho')$. Moreover, if $r_1=P+\delta$ with $\delta \to 0$, we get $\langle \bar{C}\rangle  \simeq ( 1/p'_{eq,1}+ 1/p'_{eq,2}+2\ln(p'_{eq,1}/p'_{eq,2}))\delta \eta$, thus we get $\langle \bar{C}\rangle$ negative or positive depending on the sign of $\delta$.
\end{proof}

In general, for $\rho$ such that $\Delta(\rho)=\rho_\beta$, e.g., from Eq.~\eqref{eq. b2}, % and~\eqref{eq. bar C 2}
we get
\begin{equation}\label{eq. err}
\beta(\langle w\rangle-\Delta F) = C_{\bar{\Delta}}(\rho') - C_{\Delta}(\rho) + S(\bar{\Delta}(\rho')||\rho'_\beta)\,,
\end{equation}
extending the result achieved in Ref.~\cite{Francica19} for the irreversible work to states $\rho\neq \Delta(\rho)=\rho_\beta$. Thus, the bounds in Eqs.~\eqref{eq. bound 1} and~\eqref{eq. bound 2} can be also easily derived from Eq.~\eqref{eq. err}. %, so that they hold even if $\bar{R}_n\neq U_{0,\tau}R_nU_{0,\tau}^\dagger$.
In particular, the right side of Eq.~\eqref{eq. err} corresponds to the lower bound for $\beta(\langle w\rangle-\Delta F)$ recently derived in Ref.~\cite{rodrigues23}. Thus, we note that, when the quantum system is thermally isolated as in our case, the bound of Ref.~\cite{rodrigues23} reduces to Eq.~\eqref{eq. err} and gives the exact value of the average work. %Of course, other bounds can be trivially derived from Eq.~\eqref{eq. err}, but they are not related to a fluctuation theorem.

\begin{figure}
[t!]
\centering
\includegraphics[width=0.9\columnwidth]{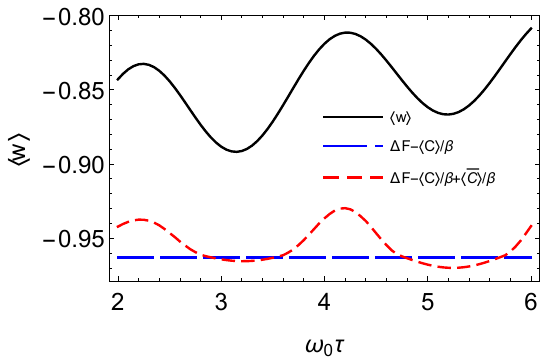}
\caption{The average work $\langle w\rangle$ in unit of $\beta$, and the two bounds coming from the fluctuation theorem. We put $\omega_\tau=2\omega_0$, $c=\sqrt{p_0(1-p_0)}/2$, with $p_0=e^{\beta \omega_0}/(e^{\beta \omega_0}+e^{-\beta \omega_0})$, $\beta \omega_0=1$ and a value of $\bar c$ that minimizes the relative entropy $S(\rho'||\bar\rho)$.
}
\label{fig:plot}
\end{figure}

To further characterize our results, e.g., the two bounds in Eqs.~\eqref{eq. bound 1} and~\eqref{eq. bound 2} deriving from the detailed fluctuation relation in Eq.~\eqref{eq. FT q}, we can focus on a physical example experimentally studied in Ref.~\cite{batalhao15}, which is a qubit with Hamiltonian $H(t) = \omega(t) (\sigma^x \cos \varphi(t)+\sigma^y \sin \varphi(t))$, where $\varphi(t) = \frac{\pi t}{2\tau}$, $\omega(t) = \omega_0 (1-t/\tau)+\omega_\tau t/\tau$ and $\sigma^x$, $\sigma^y$ and $\sigma^z$ are the Pauli matrices. For studying the effect of the initial coherence we take the initial density matrices $\rho= \rho_\beta + c \sigma^z$ and $\bar \rho= \rho'_\beta + \bar c \sigma^z$. Then, in Fig.~\ref{fig:plot} we plot the average work and the lower bounds in the function of the duration time $\tau$, illustrating the previous discussion with a physical process. We note that although $\bar c$ is chosen such that the state $\bar \rho$ minimizes the relative entropy $S(\rho'||\bar\rho)$, the bound in Eq.~\eqref{eq. bound 2} is not always tighter than the bound in Eq.~\eqref{eq. bound 1}, which does not depend on the time-evolution.

\section{Conclusions}

Tasaki-Crooks fluctuation theorem still holds in the presence of initial quantum coherence if the populations of the initial states of the forward and backward processes are thermal and we consider the two-projective-measurement scheme to infer the work statistics. However, this scheme becomes invasive when there is initial quantum coherence, e.g., it gives a different value for the average work. Here, going beyond the two-projective-measurement scheme result, we proved that a more general detailed fluctuation theorem can be formulated by considering a quasiprobability distribution of work, which implies the two-projective-measurement scheme Tasaki-Crooks fluctuation theorem.
Furthermore, it also implies two different integral fluctuation relations for the forward process, bounding the average work done in this process. Thus, our results aim to clarify the effects of quantum coherence to the Tasaki-Crooks fluctuation theorem, showing how the latter is affected by the initial quantum coherence and changes form.  % when the work is not inferred through the two-projective-measurement scheme but it is represented by a quasiprobability distribution.}
Furthermore, it is worth noting that this general detailed fluctuation relation involves a backward quasiprobability distribution that does not necessarily satisfy the reproduction of the two-projective-measurement scheme. Thus, the contextuality of the backward protocol, which is represented by suitable quasiprobability distributions reproducing the two-projective-measurement statistics for incoherent initial states, is not constrained by the contextuality of the forward one.
In conclusion, we hope that our results can help to clarify the effects of the presence of initial quantum coherence in work fluctuation theorems, which are not taken into account by the two-projective-measurement scheme.

\subsection*{Acknowledgements}
The authors acknowledge financial support from the project BIRD 2021 "Correlations, dynamics and topology in long-range quantum systems" of the Department of Physics and Astronomy, University of Padova, from the EuropeanUnion-NextGenerationEU within the National Center for HPC, Big Data and Quantum Computing (Project No. CN00000013, CN1 Spoke 10 Quantum Computing) and  from the Project "Frontiere Quantistiche" (Dipartimenti di Eccellenza) of the Italian Ministry for Universities and Research.

\appendix

\section{Quasiprobabilities}\label{appendix}
We recall our notion of quasiprobability introduced in Ref.~\cite{Francica222}.
In general, events are represented as effects, which are the positive operators which can occur in the range of a positive operator valued measurement, i.e., an effect is a Hermitian operator $E$ acting on the Hilbert space $\mathcal H$ such that $0\leq E \leq I$.
For a single event, the generalized probability measures on the set of effects are functions $E\mapsto v(E)$ with the properties
\begin{eqnarray}
\label{P1}&& 0\leq v(E) \leq 1\,,\\
\label{P2}&& v(I)=1\,,\\
\label{P3}&& v(E+F+\cdots)=v(E)+v(F)+\cdots
\end{eqnarray}
where in Eq.~\eqref{P3}, $E+F+\cdots \leq I$. A theorem~\cite{bush03} states that, if Eqs.~\eqref{P1}-\eqref{P3} are satisfied, then the probability corresponding to the event represented by $E$ is $v(E)=\Tr{E \rho}$ for some density matrix $\rho$.
For two events, we can define a function $v(E,F)$ with the properties~\cite{Francica222}
\begin{eqnarray}
\label{Q1}&& v(E,F)\in \R \,,\\
\label{Q2}&& v(I,E)=v(E,I)=v(E)\,,\\
\nonumber && v(E+F+\cdots,G)=v(E,G)+v(F,G)+\cdots\,,\\
\label{Q3}&& v(G,E+F+\cdots)=v(G,E)+v(G,F)+\cdots
\end{eqnarray}
where in Eq.~\eqref{Q3}, $E+F+\cdots \leq I$.
If Eqs.~\eqref{Q1}-\eqref{Q3} are satisfied, and if $v(E,F)$ is sequentially continuous in its arguments, then the quasiprobability corresponding to the events $E\land F$ is a bilinear function, in detail it is $v(E,F)=\text{Re}\Tr{E F \rho}$ for some density matrix $\rho$ (see Ref.~\cite{Francica222} for details).
Similarly, for three events, we define a quasiprobability $v(E,F,G)$ with the properties
\begin{eqnarray}
\label{Qq1}&& v(E,F,G)\in \R \,,\\
\label{Qq2}&& v(I,E,F)=v(E,I,F)=v(E,F,I)=v(E,F)\,,\\
\nonumber && v(E+F+\cdots,G,H)=v(E,G,H)+v(F,G,H)+\cdots\,,\\
\nonumber && v(G,E+F+\cdots,H)=v(G,E,H)+v(G,F,H)+\cdots\,,\\
\label{Qq3}&& v(G,H,E+F+\cdots)=v(G,H,E)+v(G,H,F)+\cdots\,,
\end{eqnarray}
and in general, for an arbitrary number of events, we define a quasiprobability $v(E,F,\cdots)$ with the properties
\begin{eqnarray}
\label{Q1M}&& v(E,F,\cdots)\in \R \,,\\
\label{Q2M}&& v(I,E,F,\cdots)=v(E,I,F,\cdots)=\cdots=v(E,F,\cdots)\,,\\
\nonumber && v(E+F+\cdots,G,\cdots)=v(E,G,\cdots)+v(F,G,\cdots)+\cdots\,,\\
\label{Q3M}&& \cdots
\end{eqnarray}
where in Eqs.~\eqref{Qq3} and~\eqref{Q3M}, $E+F+\cdots \leq I$. Analogously, if Eqs.~\eqref{Q1M}-\eqref{Q3M} are satisfied, and if $v(E,F,\cdots)$ is sequentially continuous in its arguments, then the joint quasiprobability corresponding to the events $E\land F\land \cdots$ can be expressed as an arbitrary affine combination of $\text{Re} \Tr{X_i \rho}$ where the operators $X_i$ are all the possible products of the effects, e.g., for two events we can consider only the product $X_1=EF$, since $X_2=FE$ gives the same quasiprobability; for three events, we can consider the three products $X_1=EFG$, $X_2=FEG$ and $X_3=EGF$, and so on.
Basically, the quasiprobability is not fixed for more than two events since the proposition $E\land F\land \cdots$ is not well defined.
In particular, for more than two events, the quasiprobability depends on how the events are grouped together.
For instance, for three events, a proposition $E\land F \land G$ can be decomposed in three different ways, which are $E\land F$, $F \land G$, or $F\land E$, $ E\land G$, or $E\land G$, $ G\land F$, then there is a one to one correspondence between the quasiprobabilities $\text{Re} \Tr{X_i \rho}$ and the different decompositions. It is straightforward to see that this correspondence holds also for an arbitrary number of events, thus we can associate the quasiprobability  $v(E,F,G,\ldots)=\text{Re} \Tr{EFG\cdots \rho}$ to the definite decomposition $E\land F$, $F\land G$, $G\land \cdots$.

\end{document}